\documentclass[epj, nopacs]{svjour}
\usepackage{graphics}
\begin{document}
\title{Measurement of the Spin Rotation Parameter $A$
in the Elastic Pion-proton Scattering at 1.43~GeV/c}
\author{I.G.~Alekseev\inst{1}, N.A.~Bazhanov\inst{2}, P.E.~Budkovsky\inst{1},
E.I.~Bunyatova\inst{2}, V.P.~Kanavets\inst{1}, A.I.~Kovalev\inst{3},
L.I.~Koroleva\inst{1}, S.P.~Kruglov\inst{3}, B.V.~Morozov\inst{1},
V.M.~Nesterov\inst{1}, D.V.~Novinsky\inst{3}, V.V.~Ryltsov\inst{1},
V.A.~Shchedrov\inst{3}, A.D.~Sulimov\inst{1}, V.V.~Sumachev\inst{3},
D.N.~Svirida\inst{1}, V.Yu.~Trautman\inst{3} \and V.V.~Zhurkin\inst{1}
}                     % Do not remove
\institute{Institute for Theoretical and Experimental Physics, Moscow, 117218, Russia \and
Joint Institute for Nuclear Research, Dubna, Moscow area, 141980, Russia \and
Petersburg Nuclear Physics Institute, Gatchina, Leningrad district, 188300, Russia}
\date{Received: date / Revised version: date}
\abstract{
The ITEP-PNPI collaboration presents new results of the measurements of the spin
rotation parameter $A$ in the elastic scattering of negative pions on protons
at $P_{beam}=1.43$~GeV/c. The results are compared to the predictions of several
partial wave analyses. The experiment was performed at the ITEP proton synchrotron,
Moscow.
%
%\PACS{
%      {PACS-key}{discribing text of that key}   \and
%      {PACS-key}{discribing text of that key}
%     } % end of PACS codes
} %end of abstract
\authorrunning{I.G.~Alekseev et al.}
\titlerunning{Measurement of the Spin Rotation Parameter $A$
at 1.43~GeV/c}
\maketitle
\section{Introduction}
\label{intro}
The present experiment is the last in the series of spin rotation 
parameter measurements performed by the ITEP-PNPI collaboration
during the last decade~\cite{c:1}. The main goal of these studies
was to enrich the experimental database of partial wave analyses
(PWA) with qualitatively new information on the spin rotation parameters,
which have never been measured in the incident momentum interval
under consideration. The momentum range (0.8--2.1)~GeV/c available
at the ITEP beam-line is very important for the baryon spectroscopy
because it contains nearly 65\% of the known light quark resonances.
There are three clusters of resonances in this region corresponding
to the peaks in the pion-proton total elastic cross-section.
The current interest to the light baryon spectroscopy is enhanced
by several experimental observations which do not well fit the constituent
quark models. As examples one can mention the existence of resonance 
clusters with masses 1.7 and 1.9~(GeV/c)$^2$, the presence of the
negative parity resonances in the cluster at $\sqrt{s}=1.9$~(GeV/c)$^2$,
indications for the parity doublets in both mentioned clusters and
"missing resonances" problem near $\sqrt{s}=2.0$~(GeV/c)$^2$.
At the same time the current status of the experimental light baryon
spectroscopy is far from satisfactory. 

Partial wave analysis is the most powerful tool of the baryon spectroscopy.
Yet the data on the resonances presented even in the latest versions
of RPP~\cite{c:RPP} are based mainly on the two PWA: KH80~\cite{c:KH80}
and CMB~\cite{c:CMB}, both performed more than two decades ago.
However more recent analysis by VPI group~\cite{c:VPI} did not
reveal a certain number of resonances: D$_{13}$(1700), S$_{31}$(1900), 
P$_{33}$(1920), D$_{33}$(1940). In the recent years GWU-VPI group
continued the PWA development~\cite{c:GWU}.

PWA solution two-fold ambiguities lead to the significant discrepancies in
the predictions of various analyses for the spin rotation parameters
in certain kinematic regions. Since the measurement of the spin
rotation parameters is the only source of the experimental information
on the relative phase of the transverse scattering amplitude, it
appears to be an unavoidable step to the unambiguous reconstruction
of the pion-proton elastic scattering amplitude.

The present experiment is performed at the c.m. energy 1.9~GeV/c$^2$
corresponding to the I=3/2 resonance cluster with complicated structure. 

\section{Formalism}
\label{form}
The meaning of the $A$ and $R$ spin rotation measurement becomes clear
from the expressions below, where the observables are expressed in terms
of the transverse amplitudes $f^+$ and $f^-$:
\begin{eqnarray}
\sigma & = & |f^+|^2 + |f^-|^2 \nonumber \\
P \cdot \sigma & = & |f^+|^2 - |f^-|^2  \label{e:1} \\
A \cdot \sigma & = & Re(f^+f^{-*})\cdot \sin(\theta_{cm}-\theta_{lab}) - \nonumber \\
&& -Im(f^+f^{-*})\cdot \cos(\theta_{cm}-\theta_{lab}) \nonumber \\
R \cdot \sigma & = & Re(f^+f^{-*})\cdot \cos(\theta_{cm}-\theta_{lab}) + \nonumber \\
&& +Im(f^+f^{-*})\cdot \sin(\theta_{cm}-\theta_{lab})\;, \nonumber 
\end{eqnarray}
the polarization parameters obeying the relation:
\begin{equation}
P^2+A^2+R^2=1 \label{e:2}
\end{equation}
In turn the transverse amplitudes correspond to the amplitudes of the 
scattering matrix $M=g+ih(\vec\sigma\cdot\vec n)$ by simple relations:
$f^+=g+ih$, $f^-=g-ih$. It follows from (\ref{e:1}), (\ref{e:2}) that
differential cross-section and normal polarization measurements allow to reconstruct
only the {\em absolute values} of the transverse amplitudes, while
their relative phase may be obtained only from the spin rotation parameter 
measurement.

\section{Experiment layout}
\label{layout}
\begin{figure}
\resizebox{0.48\textwidth}{!}{%
  \includegraphics{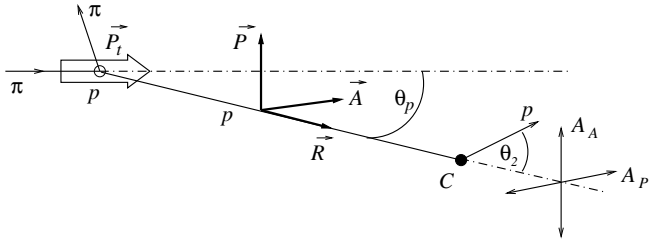}
}
\caption{The $A$ experiment idea.}
\label{f:1}
\end{figure}

Figure~\ref{f:1} shows the idea of the experiment. Two components $\vec P$
and $\vec A$ of the recoil
proton polarization are analyzed by its scattering on a carbon filter $C$,
while the proton target polarization is measured by means of NMR. In the
present experiment so-called $A$-geometry is used, i.e. the target
polarization vector $\vec P_t$ is collinear to the pion beam direction. In this case
the transverse component of the recoil proton polarization in the scattering
plane is $\vec A = P_t\cdot A$, where $P_t$ is the value of the target polarization,
$A$ -- spin rotation parameter. The component perpendicular to the scattering 
plane is equal to the normal polarization $P$. Horizontal (vertical)
component of the recoil proton polarization is determined by the measurement
of the vertical $A_A$ (horizontal $A_P$) asymmetry of the scattering on carbon.

The setup is located on the universal two-focus beam-line of the ITEP
synchrotron. The beam-line can provide pions of both signs and/or
protons in the momentum range (0.8--2.1)~GeV/c with the resolution
$\Delta p/p=\pm1.8$\%. The size of the beam spot in the target region
is close to 30~mm FWHM in both directions.

\begin{figure}
\resizebox{0.48\textwidth}{!}{%
  \includegraphics{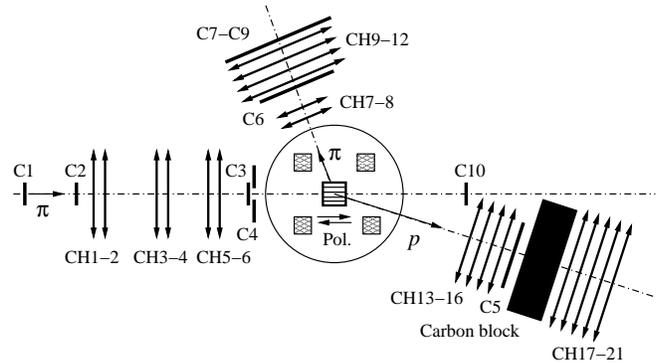}
}
\caption{Experiment layout.}
\label{f:2}
\end{figure}

The basic elements of the setup (fig.~\ref{f:2}) are: the longitudinally 
polarized proton target inside the superconducting solenoid, thick block
carbon polarimeter, sets of the wire chambers for the tracking of
the incident and scattered particles and the TOF system for the
beam particle identification. 

Evaporation type $^3$He cryostat
maintains 0.55~K in the 20~cm$^3$ target container with propanediole 
(C$_3$H$_8$O$_2$ doped with Cr$^V$ complexes), while the proton 
polarization is achieved by the dynamic nuclear orientation in the
2.5~T magnetic field of a Helmholtz pair of superconducting coils.
The polarization value is (70--80)\% with the measurement uncertainty 1.5\%,
its sign reversed every 12~hours to suppress the false asymmetries.

The block carbon polarimeter has the thickness of \linebreak 
36.5~g/cm$^2$. Since
the uncertainty of the analyzing power directly contributes to the
systematic error of the measured $A$-parameter, the polarimeter
was calibrated in advance using the polarized beam of protons
elastically scattered from the internal accelerator targets (polyethylene
or carbon). The carbon analyzing power was measured as a function of
proton momentum and scattering angle with the $\approx$4\% error
of its average in the angular interval of interest. 

\section{Data processing}
\label{dataproc}
The data processing included the following main steps:
\begin{itemize}
\item selection of the elastic scattering events on the free protons of the polarized 
target;
\item selection of the recoil proton on carbon scattering events in the 
angular interval (3-20)$^o$;
\item application of the maximum likelihood method to determine $A$ and $P$
parameters.
\end{itemize}

\begin{figure}
\resizebox{0.48\textwidth}{!}{%
  \includegraphics{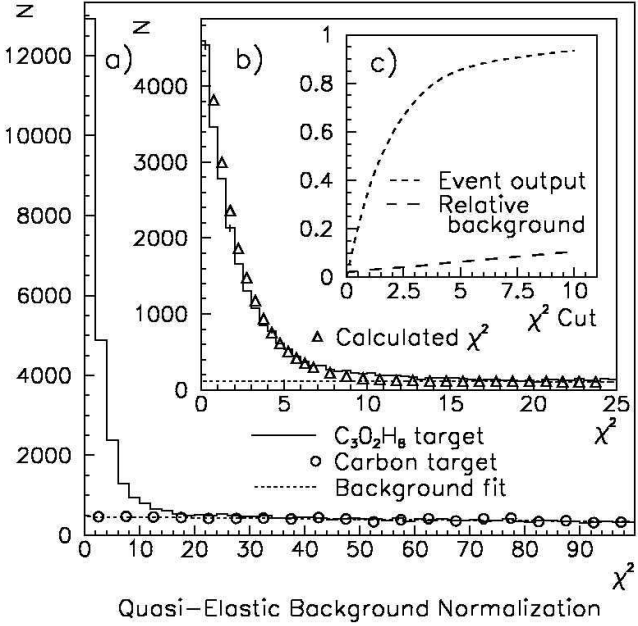}
}
\caption{$\chi^2$ distribution for the elastic and quasi-elastic events;
insertion b): the same but with smaller vertical scale, triangles designate
the theoretical $\chi^2$ distribution; insertion c): fractions of the background
(long dashes) and of the event throughput (short dashes) dependent 
on $\chi^2_{cut}$ value.}
\label{f:3}
\end{figure}

Elastic scattering events were selected by the angular correlations between
the scattered pion and the recoil proton (polar angles and co-planarity).
United $\chi^2$ criteria which accounts for both correlations was:
$$
\chi^2 = (\Delta \theta / \sigma_{\theta})^2 + (\Delta \varphi / \sigma_{\varphi})^2\;,
$$
where $\Delta \varphi$ and $\Delta \theta$ are the deviations from the
elastic kinematics, $\sigma_{\varphi}$ and $\sigma_{\theta}$ -- the widths
of the corresponding distributions. Fig.~\ref{f:3} shows the $\chi^2$- 
distributions for the events from the polarized target and its carbon
replacement. One can see that the quasi-elastic scattering background
under the elastic peak can be  properly estimated by the linear extrapolation
of the $\chi^2$-distribution from the background region. The shape
of the elastic histogram is consistent with the theoretical
$\chi^2$-distribution for 2 degrees of freedom. Elastic events were selected
by applying the cut for their $\chi^2$ distribution. The background
fraction and the loss of the elastic events depend on the $\chi^2_{cut}$
cut value. Both dependencies are shown in the top right insertion
in fig.~\ref{f:3}. The cut at $\chi^2_{cut}=8$ was chosen in the
present experiment, resulting in the quasi-elastic background fraction
of 9\% and $\approx$10\% loss of the events.

At the next step of the data processing the vertex of the proton-carbon
interaction was determined together with the scattering angle. Events in
the angular region of high pC analyzing power (3--20)$^o$ were  selected
for further processing.

Polarization parameters $A$ and $P$ were chosen as independent
variables for the maximum likelihood method to describe the
azimuthal asymmetries of the proton-carbon scattering.
The statistical material was divided into 3 subintervals
in c.m. scattering angle. Average parameters over the whole
angular range of the measurement were also calculated.
The maximum likelihood function $L(\vec P, \vec B)$ was taken in the 
following form:
\begin{eqnarray}
\ln L(\vec P, \vec B) = \hspace{6cm} \\
\hfil = \sum^N_{i=1} k_i \ln \left ( 
1 + a_{pC}(\theta_{2i}, T_i) \cdot \vec n_2 [(1-f)\vec P'_i + f \vec B'_i]
\right )\:, \nonumber
\end{eqnarray}
where
\begin{itemize}
\item[] $\vec P$ and $\vec P'$ -- vectors of the polarization for the
recoil proton from the elastic scattering events:
in the polarized target and in the carbon block with the account of the
spin motion in the magnetic field of the target;
\item[] $\vec B$ and $\vec B'$ -- similar values for the protons, 
quasi-elastically scattered in the polarized target;
\item[] $\vec n_2$ -- unit vector to the second scattering plane;
\item[] $f$ -- fraction of the quasi-elastic background;
\item[] $a_{pC}(\theta_2, T)$ -- analyzing power of the pC scattering
as a function of the scattering angle and of the proton kinetic energy
at the pC-vertex;
\item[] $k=\eta^{-1}(\varphi_2)$, where $\eta(\varphi_2)$ is the average
polarimeter efficiency as a function of the second scattering azimuthal
angle.
\end{itemize}
\noindent
The vector $\vec P$ depends on the first scattering angle, on the target
polarization and on the values of $A$ and $P$ parameters. The vector
$\vec B$ is normal to the first scattering plane.

Using the statistical sample corresponding to an unpolarized target
the polarimeter efficiency can be expressed in the following form:
\begin{eqnarray}
\eta(\varphi_2) = \hspace{7cm} \label{e:4} \\ 
\hfil = \frac{(N^+ + N^-)(|P^+| + |P^-|)}{4(1+\cos \varphi_2 |\vec P'_N|)}
\left ( \frac{D^+(\varphi_2)}{N^+ |P^+|} + \frac{D^-(\varphi_2)}{N^- |P^-|} 
\right )\:,
\nonumber
\end{eqnarray}
where signs in the superscripts correspond to the states of the polarized 
target and
\begin{itemize}
\item[] $D^+(\varphi_2)$, $D^-(\varphi_2)$ -- densities of the azimuthal angle
distributions;
\item[] $N^+$, $N^-$ -- numbers of events;
\item[] $P^+$, $P^-$ -- values of the target polarization;
\item[] $\vec P'_N$ is the result of the transformation of the vector
$\vec P_N=(1-f)\vec P + f\vec B=[(1-f)P+fB]\cdot\vec n_1$ and $\vec n_1$ is the unit
vector normal to the first scattering plane.
\end{itemize}
For the efficiency calculations the value of $P$ parameter may be taken from
the existing experimental data or from the PWA predictions. Moreover, one
can see from (\ref{e:4}) that in the direction of the asymmetry, corresponding
to the $A$ parameter measurement (on average $\varphi_2=\pm \pi/2$) the efficiency
does not depend on the chosen value of $P$. Thus the parameter $A$ can be
determined unambiguously and with the correction account for the instrumental
asymmetry. Contrary to that, there are no tools to compensate for the
instrumental asymmetries for the $P$ parameter determination. 

In real efficiency calculations the value of $P$ parameter was taken
from the predictions of FA02 analysis~\cite{c:GWU} and averaged over
the angular acceptance of the setup $155^o<\theta_{cm}<172^o$ resulting
in the value of $<P>=-0.87$ (compare with our result in table~\ref{t:1} below).
The same efficiency function was used for all three angular subintervals.

\section{Results and systematic errors}
\label{result}
\begin{figure*}
\centerline{
 \resizebox{0.85\textwidth}{3.58cm}{%
  \begin{tabular}{ccc}
    \includegraphics{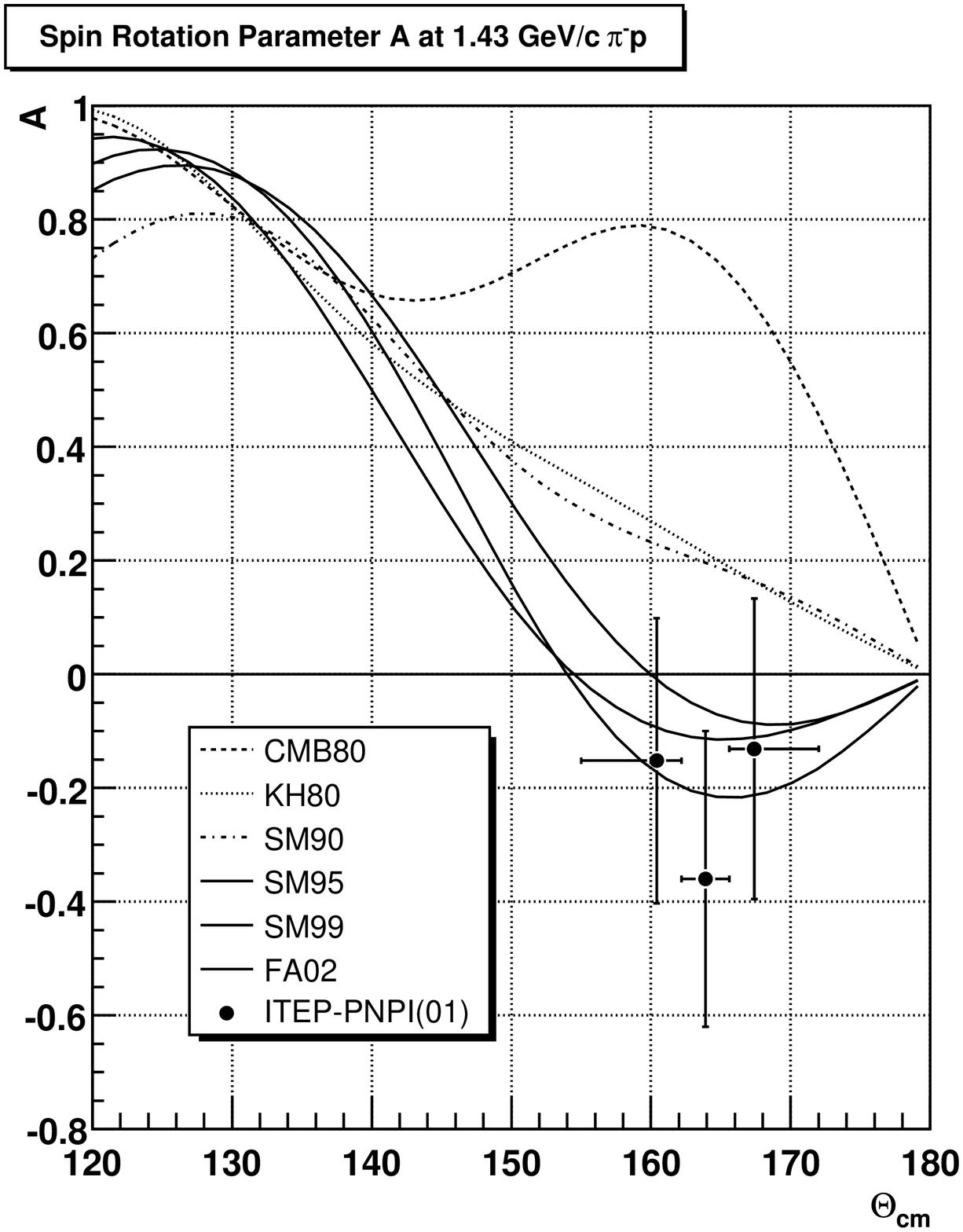} &
    \rule{3em}{0mm} &
    \includegraphics{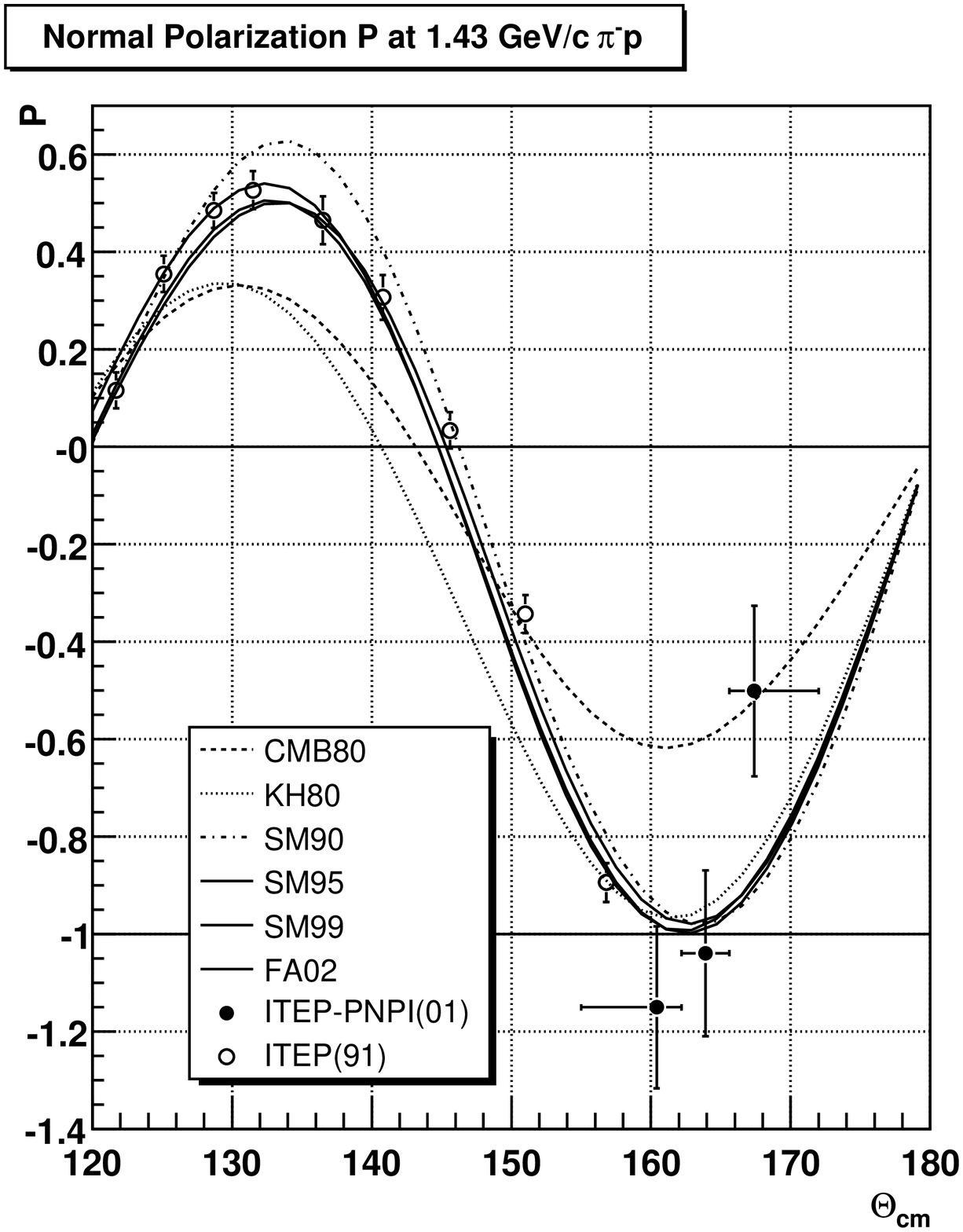} \\
  \end{tabular}
 }
}
\caption{Polarization parameters $A$ and $P$ in the $\pi^-p$ elastic scattering
at 1.43~GeV/c.}
\label{f:4}
\end{figure*}

The resulting data for $A$ and $P$ parameters are summarized in table~\ref{t:1}
and shown in figure~\ref{f:4}.
\begin{table}
\caption{Polarization parameters $A$ and $P$ in the $\pi^-p$ elastic scattering
at 1.43~GeV/c.}
\label{t:1}
\begin{tabular}{llll}
\hline\noalign{\smallskip}
\multicolumn{2}{c}{$\theta_{cm}$, deg} & $A$ & $P$  \\
range & mean & & \\
\noalign{\smallskip}\hline\noalign{\smallskip}
155--162.2   & 160.4 & $-0.152\pm0.251$ & $-1.150\pm0.166$ \\
162.2--165.6 & 163.9 & $-0.360\pm0.260$ & $-1.039\pm0.170$ \\
165.6--172.0 & 167.4 & $-0.131\pm0.264$ & $-0.501\pm0.175$ \\
\noalign{\smallskip}\hline\noalign{\smallskip}
155--172     &       & $-0.219\pm0.149$ & $-0.91\pm0.10$ \\
\noalign{\smallskip}\hline
\end{tabular}
\end{table}

To check the independence of the fitted value of $A$ parameter
on the choice of $P$ value for the efficiency calculations, it was varied
in the range $-(0.7-0.95)$ in each of the three subintervals.
The corresponding changes of $A$ parameter were within 0.01.

The most important sources of the $A$ parameter systematic errors
are the uncertainties in the polarimeter analyzing power ($\approx4$\%)
and in the measurement of the target polarization ($\approx2$\%). The
contribution of the uncertainties of the background polarization is 
negligible. This was checked by varying the value of the background
polarization from 0 to normal polarization in the elastic scattering.
The final value of the ratio of the background polarization to the
normal elastic scattering polarization was taken equal 0.7, similar
to the ratio of polarizations in the elastic and quasi-elastic
pp scattering, measured in~\cite{c:ANPC}.

The combined estimate for the systematic error of $A$ parameter 
is $\approx8$\%. 

As it was stressed above there no ways to eliminate the false
asymmetry in the direction of the $P$ parameter measurement 
($\varphi_2=0,\pi$ plane) using our experimental data. Consequently
the systematic error on $P$ cannot be reliably evaluated. Thus
the data on this parameter can be used only for qualitative
comparison with the main features of PWA solutions, for example:
strong angular dependence, angular position of the minimum
and the value at the minimum close to $-1$.

\section{Conclusion}
Measurements of the polarization parameters were performed
in the region of backward scattering where the predictions of
various PWA have maximum discrepancy. The data on $A$ parameter
agree well with SM95, SM99 and FA02 PWA solutions of the GWU-VPI
group. They contradict to the CMB80 analysis predictions while
the deviation from KH80 is three standard errors. The data on $P$
parameter is consistent with the main features of the latest
analyses of the GWU group.

The results of this experiment along with our previous measurements
of the spin rotation parameters indicate that CMB80 and KH80 analyses
do not reconstruct properly the relative phase of the transverse
amplitudes for the scattering to the backward hemisphere. This conclusion
together with the fact that the latest analysis of the GWU group 
confirm only 4-stars resonances (and only one 3-stars resonance D$_{35}$(1930))
impose serious doubts about the present-day RPP spectrum and
properties of the light baryon resonances. The development of a new
energy independent partial wave analysis or the resurrection of KH80
analysis~\cite{c:KHR} would be extremely important to establish the 
modern and reliable picture of the light quark baryon resonances.

\section{Acknowledgments}
Our thanks to professor G.H\"oehler for the interesting and fruitful
discussion. We are grateful to the staff of the ITEP accelerator for the
excellent beam quality.

The work was partially supported by Russian Fund for Basic Research
(grants 99-02-16635 and 00-15-96545) and by Russian State program
"Fundamental Nuclear Physics".

\end{document}